\documentclass{article}                    
\usepackage{subfigure} 
\usepackage{graphics,epsfig,psfig}
\usepackage{timesmt}   
\usepackage{verbatim}

\begin{document}


\title{Local and Global relations between the number of contacts and density in monodisperse sphere packs}

\author{ T. Aste, M. Saadatfar and T.J. Senden
 \\
\\ 
{\it Department of Applied Mathematics, }
\\ {\it Research School of Physical Sciences and Engineering,} 
\\ {\it The Australian National University, 0200 Australia.}}

\date{Published on: \\
{\bf J. Stat. Mech. (2006) 07010 }; \\
DOI: 10.1088/1742-5468/2006/07/P07010}

\maketitle

\begin{abstract}
The topological structure resulting from the network of contacts between grains (\emph{contact network}) is studied for large samples of monosized spheres with densities (fraction of volume occupied by the spheres) ranging from 0.59 to 0.64.
We retrieve the coordinates of each bead in the pack and we calculate the average coordination number  by using three different methods. 
We show that,  in the range of density investigated,  the coordination number is larger than 4 and it increases with the packing fraction. 
At local level we also observe a positive correlation between local packing fraction and number of neighbors.
We discover a dependence between the local densities of configurations with few neighbors in contact and the global sample-denities. 
This might indicate that local configurations with small number of neighbors are able to deform plastically when the sample is compactifying.
\end{abstract}

\section{\sc Introduction}
\label{s.I}

\begin{figure*}[t!]
\centering
\subfigure[]{\includegraphics[height=.3\textwidth]{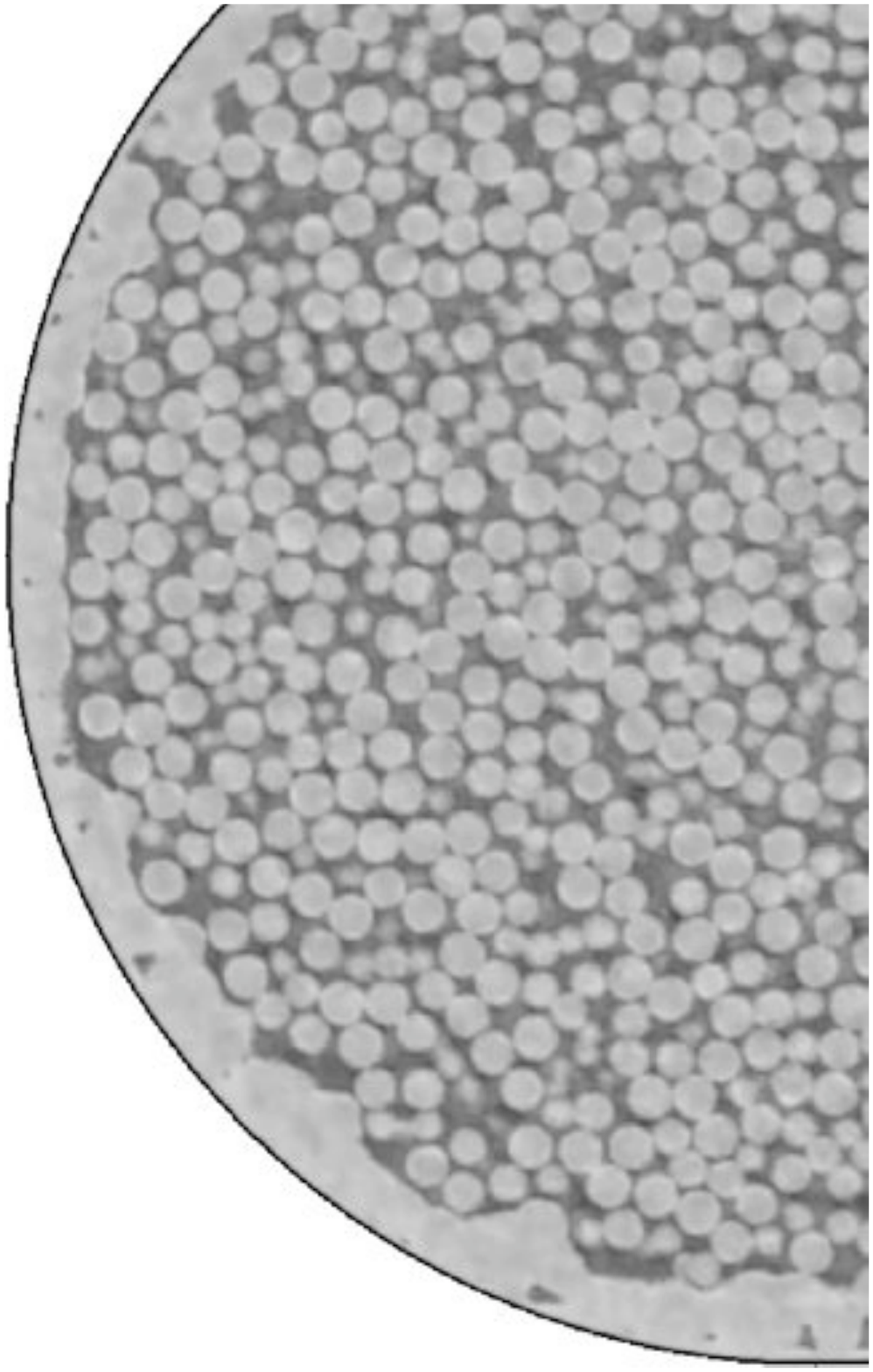}}
\subfigure[]{\includegraphics[width=.45\textwidth]{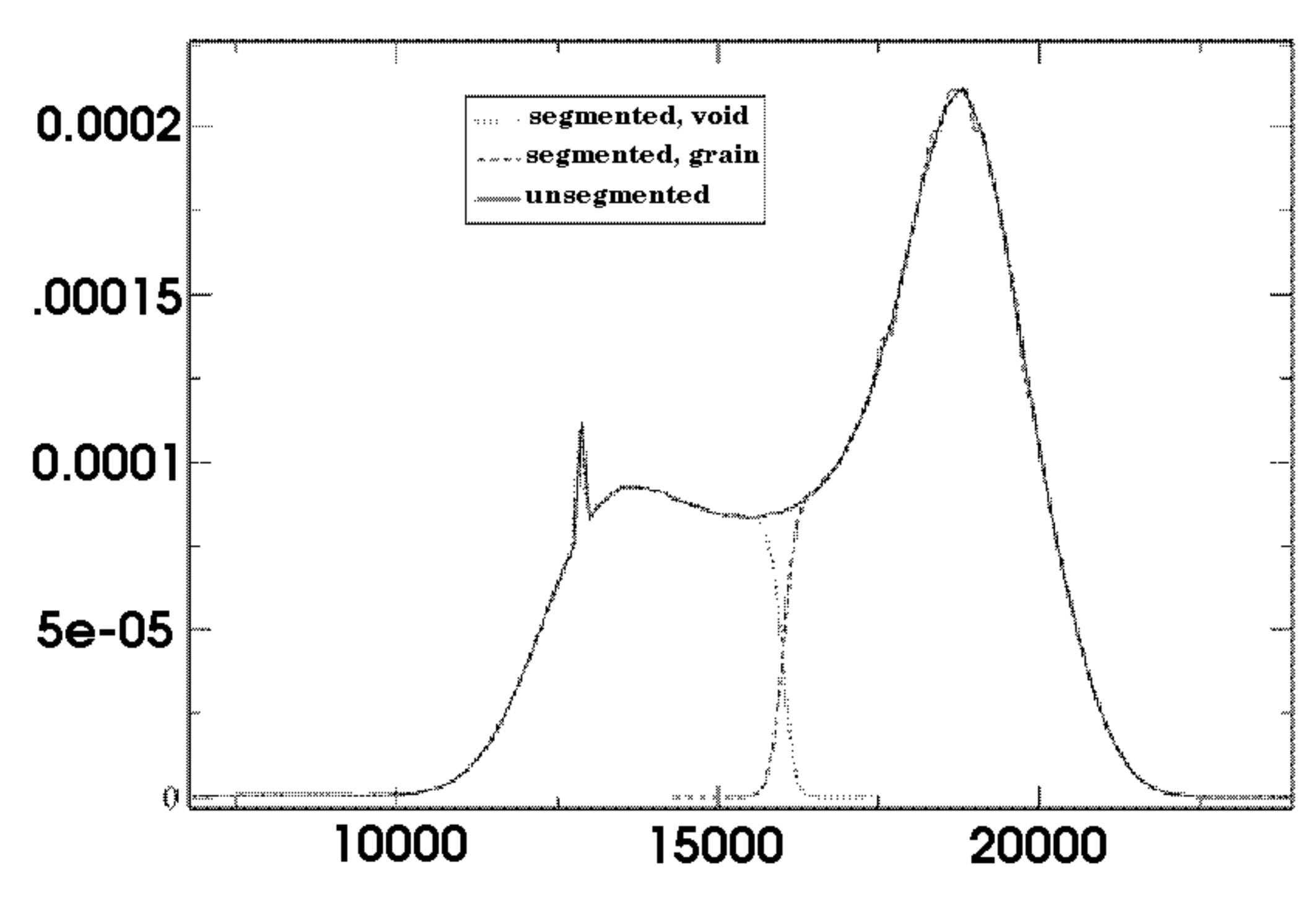}}
\subfigure[]{\includegraphics[height=.3\textwidth]{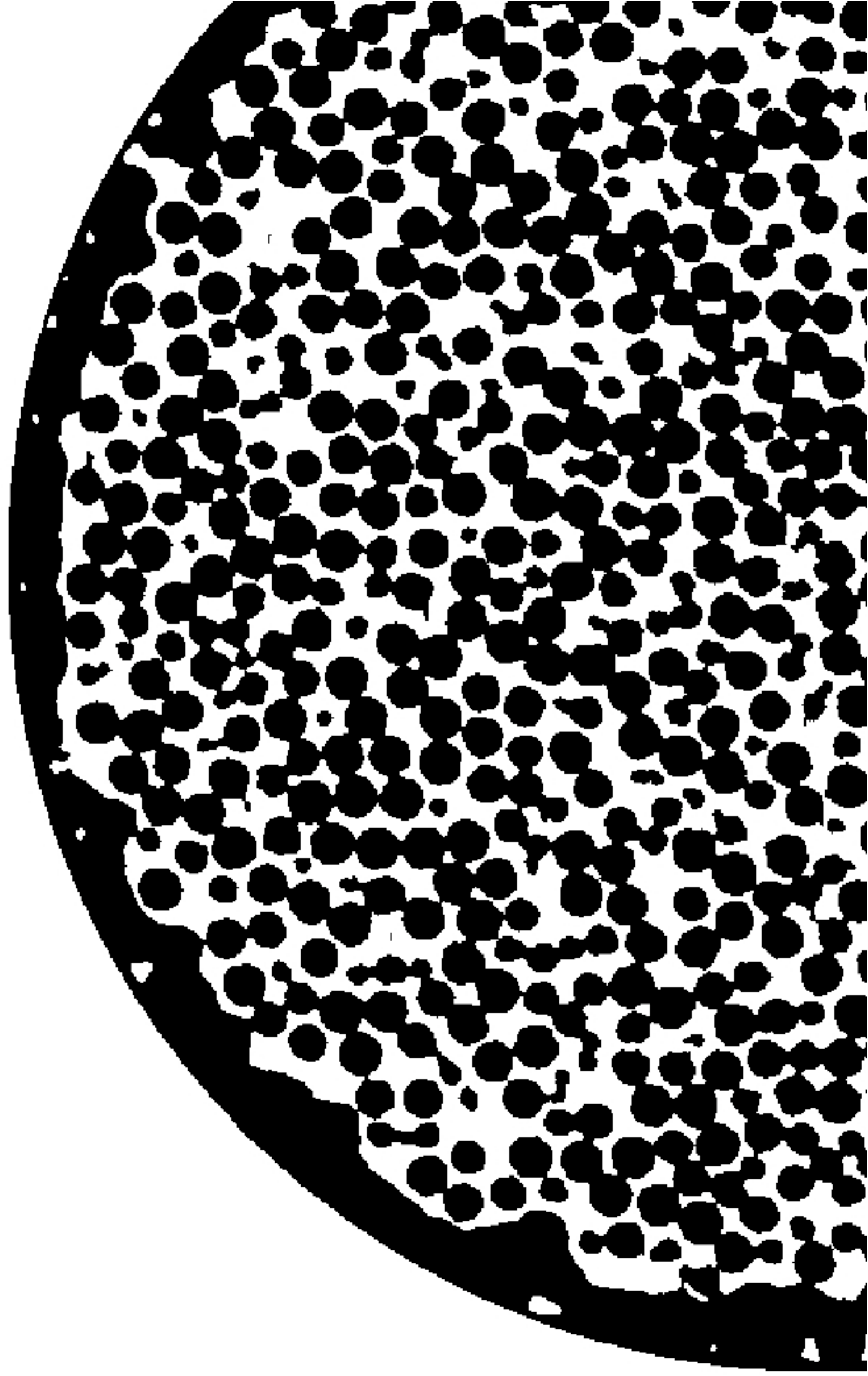}}
\caption{\footnotesize (a) Grey-scale X-ray density map 
of a slice of sample C . 
(b) Density histogram (x-axes attenuation, y-axes frequency) for the full image volume of sample C , 
and (c) the same slice 
after phase separation into pore and solid phases.}
\label{hist}
\end{figure*}

The science of granular matter has a long history with classic contributions from Faraday, Coulomb and Reynolds. 
More recently, this topic has attracted the attention of condensed matter physicists \cite{Jaeger96} and other scientists investigating the hybrid field of complex systems. 
Indeed, the contemporary science of materials and condensed-matter physics has been changing in response to a new awareness of the relevance of concepts associated with complexity. 
In this context,  granular materials have been sought as `simple' models for more complex materials and phenomena.
As such, granular materials have been typically studied either at the local level (grain scale), or as a continuous medium with specific bulk properties. 
The goal has always been to bridge these two views, in order to explain the  relationships between local structure and bulk properties.

In order to link local-structural with global, average, bulk properties  it is necessary to develop innovative ways of analyzing the system at the grain level and to explore which are the possible combinations of local configurations which generate the global packing.
Until now the empirical investigation of the geometrical structure of granular 
packings have been limited by the scarcity of accurate experimental data. Indeed, 
after the seminal works of Bernal, Mason and Scott \cite{Bernal60,Scott62}, 
it has been only very recently that the use of tomography has allowed scientists to `see' 
the three-dimensional structure of such systems and explore their geometry from the 
grain level up to the whole packing \cite{Seidler00,Sederman01,Richard03,AsteKioloa,AstePRE05,Aste05rev}.

In this paper we report and discuss results from an experimental investigation, by means 
of X-ray Computed Tomography, on very large samples of disorderly packed monosized spheres 
with densities\footnote{The density is defined as the fraction of the volume occupied by 
the balls divided by the total volume of the region of the space considered.} ranging from 
0.58 to 0.64.  This study is the largest and the most accurate empirical analysis of 
disordered packings at the grain-scale ever performed \cite{AstePRE05}.

\section{EXPERIMENTAL APPARATUS AND ANALISYS}
Detailed descriptions of the experimental methodology and apparatus are reported in \cite{AsteKioloa,AstePRE05}. 
Let us just recall that we analyzed, by means of X-ray computed tomography, 6 samples (A, B, C, D, E, F) of monosized acrylic (polymethylmethacrylate) spheres of two 
different diameters, placed in cylindrical containers with roughened walls and with an inner diameter of $ 55\; mm$, filled to a height of $\sim 75\; mm$. 
The fraction of volume occupied by the spheres in each sample and  the number of spheres  in the central region in which the analysis is performed ($N$) are shown in Table~\ref{t.1}.

\begin{table*}[t!]
\begin{minipage}{\textwidth}
\begin{center}
\begin{tabular}{ccccccccc}

\hline
   & density & $N$ & $n_t(1.02)$ & $n_t(1.05)$ & $n_t(1.10)$ &  $n_{Deconv}$ & $n_{Geom}$\\
\hline
A  & $0.586 \pm 0.005$ & 54719  & 5.5 & 6.7 & 7.5 & 5.81 & 6.81 \\
B  & $0.596 \pm 0.006$ & 15013  & 5.9 & 6.8 & 7.7 & 5.91 & 7.14 \\
C  & $0.619 \pm 0.005$ & 91984  & 6.4 & 7.5 & 8.4 & 6.77 & 7.63 \\
D  & $0.626 \pm 0.008$ & 15725  & 6.0 & 7.5 & 8.4 & 6.78 & 7.73 \\
E  & $0.630 \pm 0.010$ & 15852  & 6.3 & 7.6 & 8.6 & 6.95 & 8.14 \\
F  & $0.640 \pm 0.005$ & 16247  & 6.9 & 7.9 & 8.9 & 6.97 & 8.23 \\
\hline

\end{tabular}
\end{center}
\caption{\small Sample density and their interval of variations ($\pm$) 
within each sample; number of spheres in the central region in which we 
perform the analysis ($N$); average number of neighbors within a given threshold distance ($n_t(r/d)$) for different distances ($r/d =1.02,~1.05,~1.10$); 
average number of neighbors resulting from the deconvolution method $n_{Deconv}$;
average number of neighbors resulting from the geometry-based method $n_{Geom}$.}
\label{t.1}
\end{minipage}
\end{table*}

\subsection{Reconstructed X-ray density map and material phase identification (segmentation)}

Figure~\ref{hist}(a) shows an example of an X-ray density map 
for a slice within a tomogram of sample C. The tomographic image consists 
of a cubic array of unscaled density values, each corresponding 
to a finite cubic portion of the sample-volume (voxel). 

The first step of the reconstruction process is to differentiate the attenuation map into distinct pore and 
solid (grain) phases.
The density histogram (Fig.~\ref{hist}(b))
shows two distinct peaks associated with the two phases. 
The peak centered around $19500$ is associated with the 
granule phase. The lower peak around $13000$ is 
associated with the void phase. 
To distinguish between the two phases, it is sufficient to 
apply a threshold (in this case at $16000$) followed by the removal of 
isolated solid clusters to eliminate noise artifacts. 
A comparison between the grey-scale (density map) and binarised (segmented) 
image of a slice of sample C is shown in Fig.~\ref{hist}(a) and (c).

\subsection{Determining the bead centers}

Let us consider a tomographic image with $V$ voxels. 
In order to separate  
individual beads within the packing, we employ 
a convolution method which consists in moving through the (segmented) bead pack $P(\vec{r})$ a reference sphere $S(\vec{r})$  with radius smaller than the beads and measuring the overlap. 
The convolution of $P$ and $S$ is:

\begin{equation}
I(\vec{r}_j) = \sum_{i=1}^{V}S(\vec{r}_j-\vec{r}_i)\cdot P(\vec{r}_i)
\label{conv}
\end{equation}

Equation~\ref{conv} can be implemented very efficiently by applying 
the convolution theorem which allows to transform the convolution into a 
product in Fourier space.

\begin{figure}[b!]
\centering
{\includegraphics[width=0.45\textwidth]{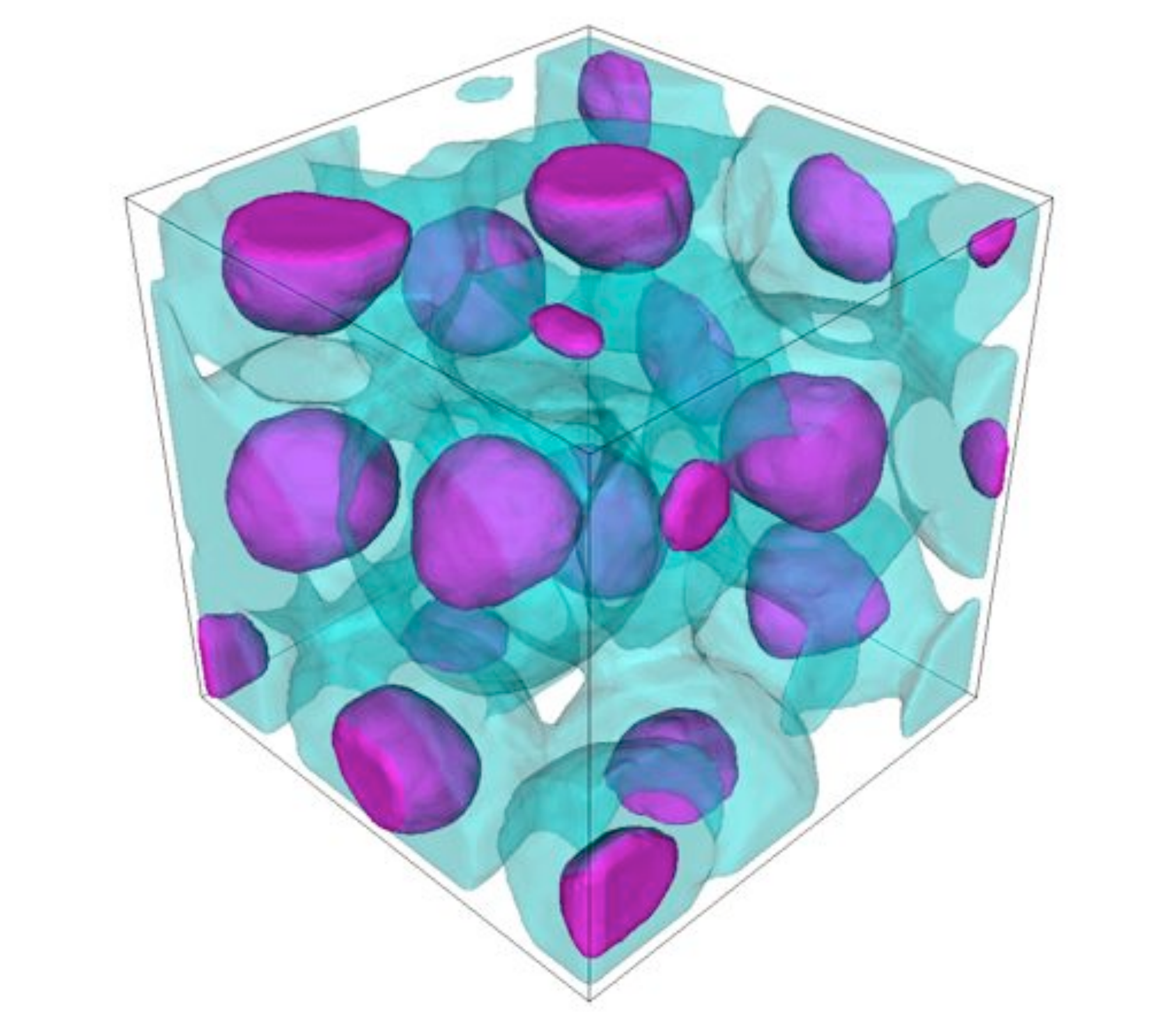}}
\caption{\footnotesize Beads (translusant) and the centeral part of each bead after applying the algorithm.}
\label{superimpose}
\end{figure}

\begin{equation}
\mathcal{F}[I] = \tilde{I}(\vec{\kappa})=\mathcal{F}[S] \cdot \mathcal{F}[P]
\label{fouriereofconv}
\end{equation}
where $\mathcal{F}$ represents the (fast) Fourier transform. 
The inverse Fourier transform of Eq.~\ref{fouriereofconv} 
results in the beads' \emph{Image function}, as expressed 
in Eq.~\ref{conv}:

\begin{equation}
I=\mathcal{F}^{-1}[\tilde{I}(\vec{\kappa})]
\label{invFT}
\end{equation}

The result of Eq.~\ref{invFT} is an intensity map of the overlap 
between the reference sphere and the beads, where the voxels 
closer to the sphere centers have higher intensities (in fact the 
peaks of the highest intensity 
represent the group of voxels around sphere centers in the original 
image). 
A threshold on the intensity map locates the groups of voxels surrounding the sphere centers. 
A visual comparison between the grains and the central part extracted by means of the present algorithm is shown in Fig.\ref{superimpose}. 
The position of the bead centers is then computed as the center of mass of each cluster weighted by the intensity of each voxel.
Below, we discuss the optimal values of the threshold  and the best choice of the reference sphere size to obtain precise estimates for the sphere centers' positions.

\subsection{Reference sphere size and intensity threshold}

\begin{figure}[b!]
\centering
\includegraphics[width=.55\textwidth]{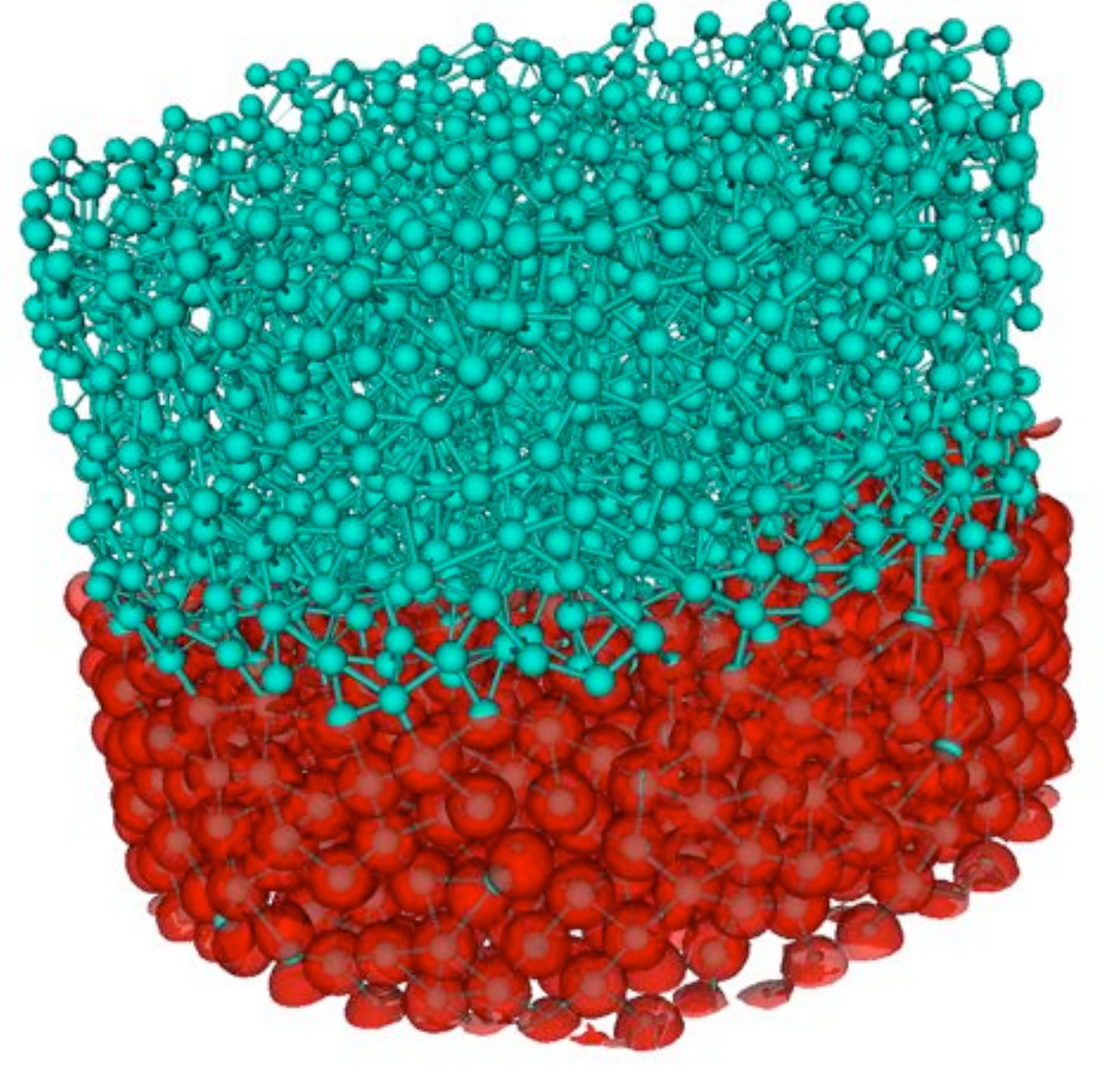}
\caption{\label{whole}  The network of connected beads in a reconstruction of sample $E$.}
\label{NetworkMeg2}
\end{figure}

The precision on the sphere centers' position depends on the spatial resolution of the 
tomograms and also on the size of the cluster surrounding the 
sphere centers after thresholding. 
The spatial resolutions of the tomograms are between one and two voxels. 
Therefore the precision on the center of mass of a cluster of $\nu$ 
voxels must be within $1/\nu$ to $2/\nu$. This suggests 
that, in order to minimize the error of the position of 
centers, we need to optimally choose the reference sphere size and 
intensity threshold so that $\nu$ will be as large as 
possible. 
The optimal values of the parameters have been chosen by
varying these two quantities and computing 
the resulting number of spheres detected in a given portion 
of the sample \cite{AstePRE05}. 
In Fig.~\ref{whole}, a reconstructed bead pack (sample E) is displayed.

\section{\sc Coordination Number}
\label{s.NNC}\label{s.5}

The average number of spheres in contact with any given sphere have been widely 
investigated in the literature of granular matter \cite{Bernal60,Scott62,Seidler00,Sederman01,ppp,Donev04c,Silb02}. 
Indeed, this is a very simple topological quantity which gives important information about the local configurations and the 
packing stability.

In this paper we discuss the construction of the `contact' networks by using three different methods.
\begin{itemize}
\item[\bf i.] A thresholding method, where all the sphere centers within a given radial distance from the centre of a sphere are considered as `in touch'. 
\item[\bf ii.] A deconvolution method which statistically distinguishes the contribution from the sphere in contact and the spheres that are close but not in touch.  
\item[\bf iii.]  A geometry-based method which employs the geometical information contained within the tomogram to verify if two grains are in contact or not.
\end{itemize}

\subsection{Thresholding and deconvolution methods}

From a pure geometrical point of view, the -empirical- definition of `spheres in contact' depends on the minimum gap within which two spheres are considered in touch. 
Indeed, the exact information on the sphere positions and their shapes and sizes is  empirically known only with some degree of incertitude.
It is quite straightforward to understand that the actual number of spheres within a radial distance $r$ from a given sphere ($n_t(r)$) must increase with the radial distance: larger is the distance, and larger is the volume of space in which the neighboring spheres are counted.

However, to  $n_t(r)$ contribute both spheres in contact and spheres that are near but do not touch.
It is also evident that the region concerned with the contribution from the spheres in contact can be only within a small region with size compatible with the sphere polydisperity.
The 6 samples analyzed in \cite{AstePRE05} have polydispersities around 2\% of their diameters, therefore it is reasonable to assume that above the radial distance $1.02d$ all the contributions to  $n_t(r)$  come only from spheres not in contact.
We can therefore (over-)estimate the number of grain in `contact' by counting all couple of spheres with centers within radial distances $r \le 1.02d$.
The results for the average numbers of spheres in `contact' at threshold distances $r/d = 1.02 $, 1.05 and 1.10 are reported in Table \ref{t.1}.

The thresholding procedure has the defect to count both the contributions from the spheres in contact and from other spheres which are near but do not touch.
The challenge is to distinguish between these two different contributions.
In \cite{AstePRE05} a deconvolution method was introduced to overcome this problem.
THe results of such an analysis are reported in the column $n_{Deconv}$ in Table \ref{t.1}.

\subsection{Geometry-based method}

Differently from the methods discussed in the previous section, this method detects the touching spheres by using directly the information from
the segmented image of granular packs.
Indeed, the segmented image provides a full 3D structural information of the bead samples and 
therefore one can directly determine whether two beads are in contact or not. 
However, this method, depends highly on the resolution of the the tomogram. Since there 
is an uncertainty\footnote{This is due to the diffraction of X-ray 
at the boundary} of $1$ to $2$ voxels in determining the grain-grain boundary, we expect to 
detect neighbors within radial distances  of $1-2$ voxels$^{1/3}$.
 This means that for beads with diameters of $25$ voxels$^{1/3}$ (B, D, E, F) \cite{AstePRE05},  the uncertainty is within radial distances of $1.04-1.08\times d$.
 Whereas an uncertainties of $1.03 - 1.07 \times d$ are expected in samples (A, C) where the diameter is 30.81 voxels$^{1/3}$ \cite{AstePRE05}. 

The results for the average number of neighbors in contact, computed by means of this geometrical-based method are reported in Table.~\ref{t.1} ($n_{Geom}$).
An example of contact network resulting from this analysis is shown in  Fig.~\ref{NetworkMeg2}.  
A comparison between these data and the results from the deconvolution method shows a very good agreement between these two techniques 
when the implicit thresholding due to the voxel-precision is considered. 
As expected, we find that the geometry-based method overestimates the number of actual contacts giving results equivalent to the tresholding methods when neighbors within radial distances between $1.05$ and $1.07$ bead diameters are counted.

\begin{figure}[t!]
\centering
\vspace{.3cm}
\includegraphics[width=.85\textwidth]{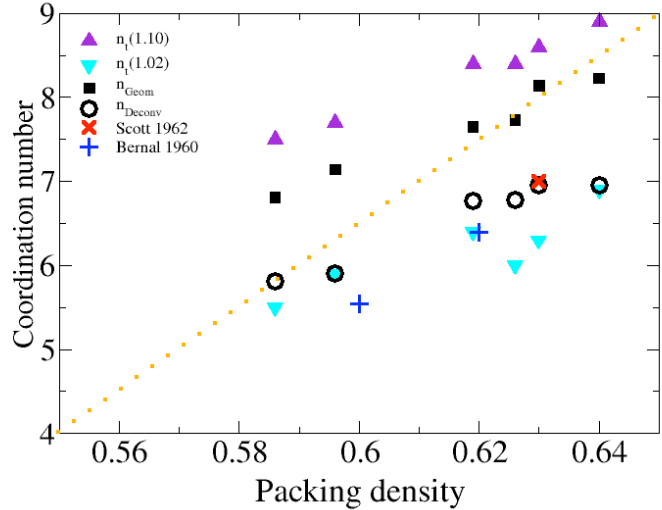}
\caption{\label{f.Nc_rho}Number of neighbors in contact vs sample's density.
The dotted line is an extrapolation to $n_c=4$ at $\rho = 0.55$.}
\label{CoordNumbPlot}
\end{figure}

In Fig.\ref{CoordNumbPlot} the values of  the average number of contacts are reported as a function of the sample densities.
From the figure it is clear that all the methods reported in this paper and also other independent experimental investigations, give  average numbers of contacts larger than 4 which  increase with the sample-density. 
Such dependence on the packing density (Fig.\ref{CoordNumbPlot}) has important  conceptual implications especially for what concerns the kind of mechanical equilibrium in the system \cite{AstePRE05,Aste05rev}.


\section{Local density fluctuations and local number of neighbors}

\begin{figure}[t!]
\centering
\vspace{.1cm}
\includegraphics[width=.95\textwidth]{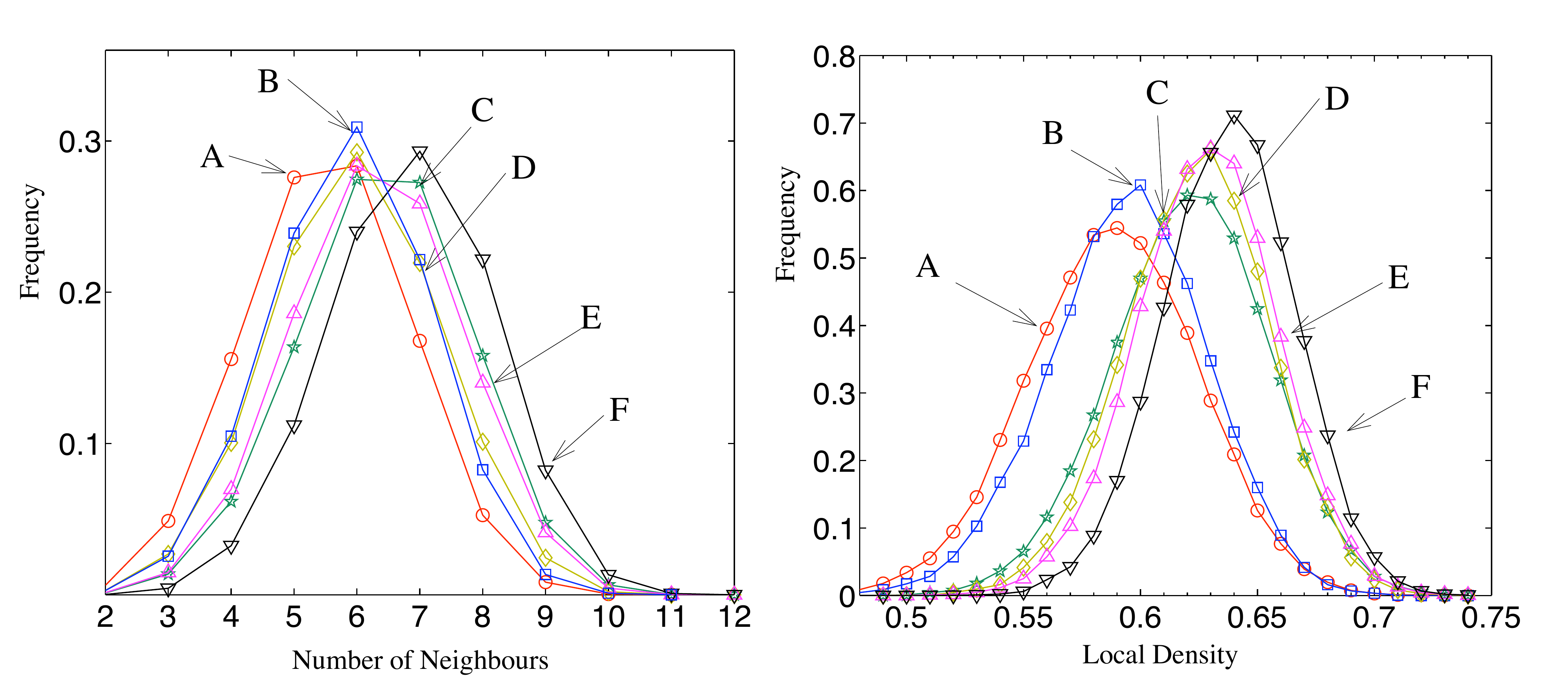}
\caption{\label{f.Nc_rho_fluct} (left) Distribution of the number of neighbors per sphere (threshold method with distance $1.02d$). 
(right) Distribution of local density (fraction of occupied Vorono\"{\i} cells). 
}
\end{figure}

In the previous section we have observed that the average number of neighbors in contact is affected by the sample-density.
Here we look at the same relation but at local level.
To this end let us first define the local density as the fraction of Vorono\"{\i} cell occupied by each sphere in the sample.
Such a quantity varies from sphere to sphere.
The distribution of local densities are reported in Fig.\ref{f.Nc_rho_fluct}(right).
One can observe that, there are no local configurations denser than 0.75 and there are very few configurations with density lower than 0.5.
The distributions have bell-like shapes with averages around the sample-densities and standard deviations between 5 and 6~\% (decreasing with increasing sample-density).
The number of neighbors also varies from sphere to sphere. 
Figure \ref{f.Nc_rho_fluct}(left) reports such a distribution for the 6 samples A-F.
As one can see, the distributions of the number of neighbors are also bell-shaped around the sample-averages and  a dependence on the sample-density is observed with denser samples showing larger connectivities.

\begin{figure}[t!]
\centering
\vspace{.3cm}
\includegraphics[width=.95\textwidth]{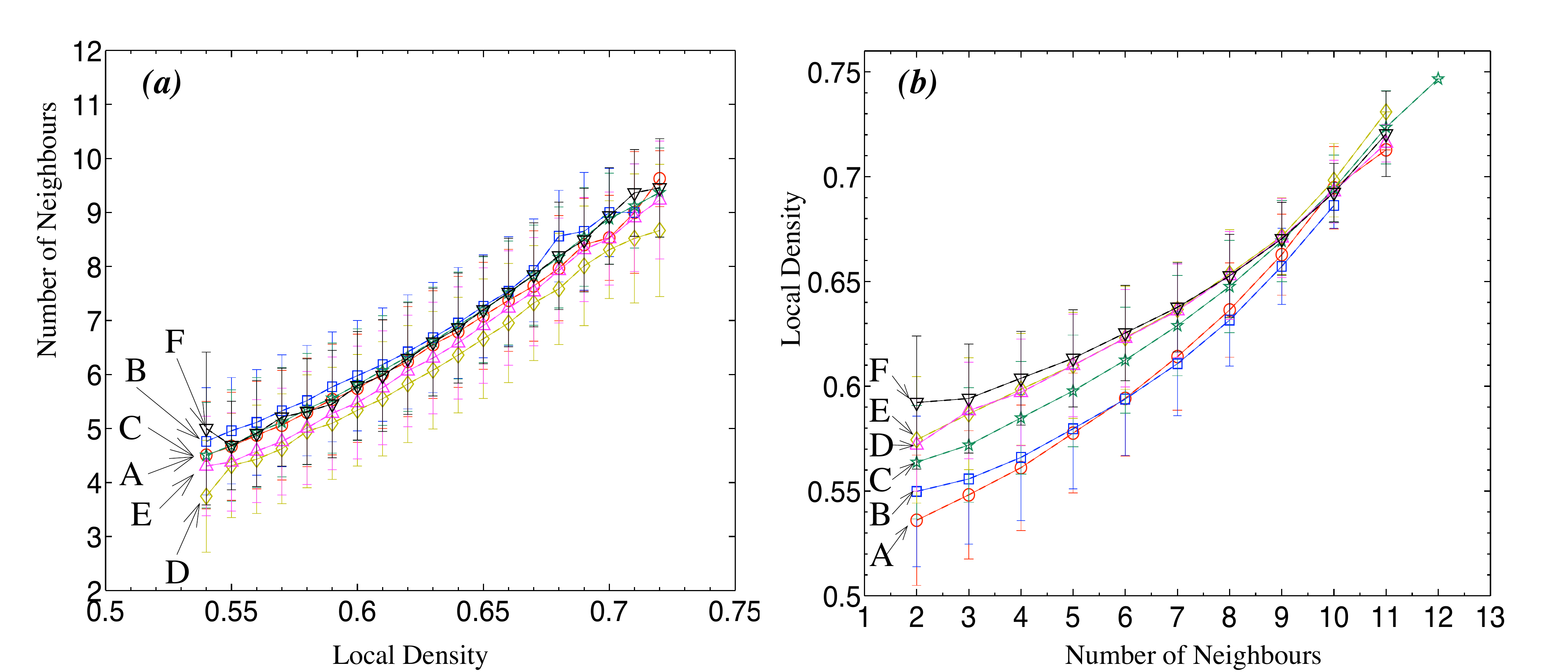}
\caption{\label{f.Nc_rho_average}
{\bf (a)} Average number of spheres counted around local configurations with given density. 
{\bf (b)} Average local density of configurations with a given number of neighbors.
In both figures the error-bars are the standard deviations.}
\end{figure}

Here we are interested in investigating the relation between the number of neighbors per sphere and the  local densities.
Such a relation is reported in the two figures Fig. \ref{f.Nc_rho_average}(a),(b) where both the average number of neighbors v.s. local density and the average local density v.s. number of neighbors are reported.
Let us first point out that, these two plots are not simply the reciprocal inverse of each other.
Indeed they correspond to two opposite gatherings.
The Fig. \ref{f.Nc_rho_average}(a)  is obtained by  counting the number of neighbors (within a threshold distance of $1.02d$) in local configurations with a given density.
Conversely, Fig. \ref{f.Nc_rho_average}(b) reports the average local densities of configurations with given number of neighbors ($1.02d$ threshold distance criteria).
In both cases the error bars correspond to the standard deviations of the respective distributions.

From these figures it is evident that there is a positive correlation between the number of neighbors and the density of local configurations.
This is not surprising, close packed configurations must have a larger number of neighbors within a given distance than loose packed configurations.
Indeed, it is known that the most compact configuration can have density 0.754 with 12 neighbors in contact. 
Whereas, on the opposite limit, Fig. \ref{f.Nc_rho_average}(b) shows that there can be some configurations with only two neighbors and local densities in a range between 0.5 to 0.62.
Beyond such rather intuitive outcomes, two interesting fact emerge from Figs. \ref{f.Nc_rho_average}: 
\begin{itemize}
\item[1)] local configurations with larger density have a larger number of neighbors, but this is essentially independent on the global sample density (see Fig. \ref{f.Nc_rho_average}(a)); \\
\item[2)] configurations with larger number of neighbors also show larger local densities but in this case a global dependence on the sample density is also observed (see Fig. \ref{f.Nc_rho_average}(b)).
\end{itemize}
Such a relation between local and global quantities is quite remarkable.
These data show that local configurations with a small number of neighbors have larger, or smaller, local densities depending on the global density of the sample.
This might suggest that the geometrical structure of sphere packings is  adapting `plastically' by expanding or  contracting some local configurations while constraining  mechanical stability.

\section{\sc Conclusion}

The study of the topological structure of the contact network at the grain level shows that the average number 
of spheres in contact increases with packing density and  lies between 5.5 and 7.5 in the range 
of densities between 0.58 and 0.64. 
An extrapolation to the random loose packing density ($\rho = 0.55$) 
suggests that at this density the system could have an average number of 4 neighbors per sphere. 
This might imply the possibility of a rigidity percolation transition taking place at the random loose packing limit \cite{Aste05rev}.
It should be however considered that the current methods to estimate the number of contacts have an excessively  large incertitude that does not allow to validate -or invalidate- the extrapolation to  $n_c = 4$ at $\rho = 0.55$.
Further studies are needed to clarify this point.
One way to reduce such incertitude on the geometrical number of neighbors is to consider in touch only grains that exchange some stress at the interface.
This can be done both by means of numerical simulations or  by direct measurements of stress in three-dimensional deformable grains. 
But, this will be the object of future research. 

At local level the we observed that there is a positive correlation between the number of neighbors and density: spheres with a larger number of neighbors occupy a larger fraction of their local Vorono\"{\i} cells.
We also observed that  configurations with small number of neighbors (2-6) are more sensitive to the global packing density.
This suggests that, while constraining global mechanical stability, the system is able to compactify by squeezing `plastically' the configurations with smaller number of neighbors.

\subsection*{Acknowledgements}
Many thanks to A. Sakellariou for the tomographic data and several discussions. This work 
was partially supported by the ARC discovery project DP0450292 and the Australian Partnership 
for Advanced Computing National Facilities (APAC).

\bibliographystyle{spmpsci}

\end{document}